\documentclass[11pt]{article}
\usepackage{amsmath,amsfonts,latexsym,graphicx,amssymb}
\usepackage{amsfonts}
\usepackage{amssymb}
\usepackage{mathrsfs}
\usepackage{amsthm}
\usepackage{graphicx}
\usepackage{amsmath}
\date{}

\newcommand{\ot}{{\,\otimes\,}}
\newcommand{{\Cd}}{{\mathbb{C}^d}}
\newcommand{{\Rn}}{{\mathbb{R}^n}}

\def\<{\langle}
\def\>{\rangle}
\newtheorem{thm}{Theorem}[section]

\newtheorem{lemma}{Lemma}
\theoremstyle{definition}
\newtheorem{ex}{Example}

\newtheorem{defn}{Definition}[section]
\newtheorem{pro}{Proposition}[section]
\newtheorem{remark}{Remark}

\begin{document}
\title{\bf On exposed positive maps: \\ Robertson and Breuer-Hall maps} \author{Dariusz
Chru\'sci\'nski \\
Institute of Physics, Nicolaus Copernicus University,\\
Grudzi\c{a}dzka 5/7, 87--100 Toru\'n, Poland}

\maketitle

\begin{abstract}

It is well known that so called Breuer-Hall positive maps used in entanglement theory are optimal.
We show that these maps possess much more subtle property --- they are exposed. As a byproduct it proves that a
Robertson map in $M_4(\mathbb{C})$ is not only extreme, which was already shown by Robertson, but also exposed.
\end{abstract}

\maketitle

\section{Exposed maps: preliminaries}

Linear positive maps \cite{Stormer}-\cite{Tomiyama} play important role in entanglement theory \cite{HHHH}: a state of a composed quantum system living in $\mathcal{H} \ot \mathcal{K}$ is separable iff $({\rm id}_\mathcal{H} \ot \phi)\rho \geq 0$ for each positive map $\phi : \mathcal{B}(\mathcal{K}) \rightarrow \mathcal{B}(\mathcal{H})$ (in this paper we consider only finite dimensional Hilbert spaces). Moreover, positive maps provide generalization of $*$-homomorphism, Jordan homomorphism and conditional expectation. Normalized positive maps define an affine mapping between sets of states of $\mathbb{C}^*$-algebras.

It is well known that a space of positive maps $\phi : \mathcal{B}(\mathcal{H}) \longrightarrow \mathcal{B}(\mathcal{K})$ is isomorphic to a space of block-positive operators in $\mathcal{B}(\mathcal{H}) \ot \mathcal{B}(\mathcal{K})$, that is, $\phi$ is positive if and only if
\begin{equation}\label{ISO}
    W_\phi = \sum_{i,j} e_{ij} \ot \phi(e_{ij})\ ,
\end{equation}
is block-positive, where $\{e_i\}$ stands for the orthonormal basis in $\mathcal{H}$. Recall, that $A$ is block-positive iff
 $   \< x \ot y |A|x \ot y \> \geq 0\,$
for all separable vectors $x \ot y \in \mathcal{H}\ot \mathcal{K}$. The inverse map is given by $\phi(X) = {\rm Tr}_\mathcal{H} (W_\phi X^{\rm t} \ot \mathbb{I}_\mathcal{K})$ for $X \in \mathcal{B}(\mathcal{H})$.  Block-positive but not positive operators are called entanglement witnesses in entanglement theory \cite{HHHH}. A density operator $\rho$ in $\mathcal{H}\ot \mathcal{K}$ is entangled iff there exists an entanglement witness $W$ such that ${\rm Tr}(\rho W) < 0$. For some recent papers discussing positive maps and entanglement witnesses see e.g. \cite{Kye,Lew,atomic,CMP,Gniewko-JPA,Augusiak}.

Let $\mathcal{P}$ denotes a convex cone of positive maps $\phi : \mathcal{B}(\mathcal{H}) \longrightarrow \mathcal{B}(\mathcal{K})$ or equivalently a convex cone of block-positive operators in $\mathcal{H}\ot \mathcal{K}$ (using (\ref{ISO}) we will identify these two cones). Let $\mathcal{P}^\circ$ denote a dual cone \cite{convex,Eom}
\begin{equation}\label{}
    \mathcal{P}^\circ = \{\  \rho \in (\mathcal{B}(\mathcal{H}) \ot \mathcal{B}(\mathcal{K}))^+\ ;\ {\rm Tr}(W\rho) \geq 0 \ , \ W \in \mathcal{P}\ \}\
\end{equation}
of separable operators (unnormalized states)  in $\mathcal{H}\ot \mathcal{K}$. Equivalently, one has
\begin{equation}\label{}
    \mathcal{P}^\circ = {\rm conv} \{\  P_x \ot P_y\ ;\ \<y|\phi(P_x)|y\> \geq 0 \ , \ \phi \in \mathcal{P}\ \}\ ,
\end{equation}
where $P_x = |x\>\<x|$ and $P_y = |y\>\<y|$. It is clear that $\mathcal{P}^{\circ\circ} = \mathcal{P}$, that is, one may consider $\mathcal{P}$ as a dual cone to the convex cone of separable operators in $\mathcal{H}\ot \mathcal{K}$. Note, that
\begin{equation}\label{}
    \<y|\phi(P_x)|y\> = \< x \ot \overline{y}|W_\phi|x \ot \overline{y}\>\ .
\end{equation}
Recall that a face of $\mathcal{P}$  is a convex subset $F \subset \mathcal{P}$ such that if the convex combination $\phi = \lambda \phi_1  + (1-\lambda)\phi_2$ of $\phi_1,\phi_2 \in \mathcal{P}$ belongs to $F$, then both $\phi_1,\phi_2 \in F$. Let $[\phi]$ denotes a ray in $\mathcal{P}$ generated by a positive map $\phi$, i.e. $[\phi] = \{ \lambda\, \phi\ ; \ \lambda >0 \}$. One says that $[\phi]$ is an extreme ray in $\mathcal{P}$  if and only if $\phi = \lambda \phi_1 + (1-\lambda)\phi_2$, with $\lambda\in (0,1)$ and $\phi_1,\phi_2 \in \mathcal{P}$, implies $\phi_1,\phi_2 \in [\phi]$. Hence, an extreme ray is 1-dimensional face of $\mathcal{P}$.

\begin{defn}
A face $F$ is exposed if there exists a supporting hyperplane
$H$ for a convex cone $\mathcal{P}$ such that $F=H \cap \mathcal{P}$.
\end{defn}
The property of `being an exposed face' may be reformulated as follows
\begin{pro}
A face $F$ of $\mathcal{P}$ is exposed iff there exists $\rho \in \mathcal{P}^0$ such that
\begin{equation}\label{}
    F = \{ \ W \in \mathcal{P}\ ; \ {\rm Tr}(\rho W)=0\ \}\ .
\end{equation}
Equivalently, there exists $x \ot y \in \mathcal{H}\ot \mathcal{K}$ such that
$    F = \{ \ \phi \in \mathcal{P}\ ; \ \<y|\phi(P_x)|y\>=0\ \}$.
\end{pro}
A ray $[\phi]$ is exposed if it defines 1-dimensional exposed face. Clearly exposed rays are extreme. We shall use the following terminology: we shall call $\phi$ an extreme (exposed)  positive map if the corresponding ray $[\phi]$ is extreme (exposed). Let us denote
by ${\rm Ext}(\mathcal{P})$ the set of extreme points and ${\rm Exp}(\mathcal{P})$ the set of exposed points of $\mathcal{P}$. Due to Straszewicz theorem \cite{convex} ${\rm Exp}(\mathcal{P})$  is a dense subset
of ${\rm Ext}(\mathcal{P})$. Thus every extreme map is the limit of some sequence of exposed
maps. It shows that each entangled state may be detected by some exposed positive map (or exposed entangled witness), i.e. $\rho$ is entangled iff there exists an exposed entangled witness $W$ such that ${\rm Tr}(\rho W) <0$. Hence, a knowledge of exposed maps (entanglement witnesses) is crucial for the full characterization of separable/entangled states of bi-partite quantum systems. For recent papers on exposed maps see e.g. \cite{Eom,Majewski,Marciniak}.

Now, if $F$ is a face of $\mathcal{P}$ then
\begin{equation}\label{dual_face}
    F' =  \{\  \rho \in \mathcal{P}^0\ ;\ {\rm Tr}(W \rho) = 0 \ , \ W \in F\ \}\ ,
\end{equation}
defines a face of $\mathcal{P}^\circ$ (one calls $F'$ a dual face of $F$). Actually, $F'$ is an exposed face. Equivalently, one has
\begin{equation}\label{dual_face}
    F' = {\rm conv} \{\  P_x \ot P_y\ ;\ \<y|\phi(P_x)|y\> = 0 \ , \ \phi \in F\ \}\ .
\end{equation}
One proves \cite{Eom}

\begin{thm} A face $F$ is exposed iff $F''=F$.
\end{thm}
Hence, $\phi$ is exposed iff the dual face $[\phi]'$ is unique, that is, if $[\psi]'=[\phi]'$, then $[\psi]=[\phi]$, or equivalently $\psi = \lambda \phi$ (with $\lambda >0$).

\section{Exposed maps: examples}

To illustrates the concept of an exposed positive map let us consider simple examples.

\begin{ex}[Transposition]
Let $\mathcal{H}=\mathcal{K}=\mathbb{C}^n$ and consider the standard transposition in $M_n(\mathbb{C})$ defined by $\tau(X) = X^{\rm t}$.
One has
\begin{equation}\label{}
    [\tau]' = {\rm conv} \{\ P_x \ot P_y\ ; \ \< y |\overline{x}\> = 0 \ \} \ ,
\end{equation}
Now, one looks for the double dual $[\tau]''$: a map $\phi \in [\tau]''$ iff $\<y|\phi(P_x)|y\>=0$ for all $x,y \in \mathbb{C}^n$ such that $\<y|\overline{x}\>=0$. Hence, $\phi(P_x) = \lambda |\overline{x}\>\<\overline{x}|$ which shows that $\phi(X)= \lambda X^{\rm t}$ and hence $\phi \in [\tau]$ which proves that $\tau$ is exposed.

\end{ex}

\begin{ex} Consider maps
\begin{equation}\label{}
    \phi_V (X) = V X V^* \ ,
\end{equation}
and
\begin{equation}\label{V-up}
    \phi^V (X) = V X^{\rm t} V^* \ ,
\end{equation}
with arbitrary $V : \mathcal{H} \rightarrow \mathcal{K}$. It is well known \cite{Yopp} that both $\phi_V$ and $\phi^V$ are extreme. Moreover, if rank of $V$ is one or $\min \{{\rm dim}\mathcal{H},{\rm dim}\mathcal{K}\}$, these maps are exposed.

\end{ex}
\begin{ex}[Reduction]
Let $\mathcal{H}=\mathcal{K}=\mathbb{C}^n$ and consider the reduction map in $M_n(\mathbb{C})$ defined by
\begin{equation}\label{}
    R_n(X) = \mathbb{I}_n\, {\rm Tr}\, X - X \ .
\end{equation}
One has
\begin{equation}\label{}
    [R_n]' = {\rm conv} \{\ P_x \ot P_y\ ; \ \<x|x\>\<y|y\> - |\<x|y\>|^2 = 0 \ \} \ ,
\end{equation}
and hence $y = \mu x$ (with $\mu \in \mathbb{C}$). Now,  $\phi \in [R_n]''$ if and only if $\<x|\phi(P_x)|x\> =0$.  Let $A$ be an arbitrary antisymmetric matrix in $M_n(\mathbb{C})$ and consider the following completely copositive map
\begin{equation}\label{}
    \phi(X) = A X^{\rm t} A^*\ .
\end{equation}
One has $\<x|\phi(P_x)|x\> = |\<x|A|\overline{x}\>|^2 = | \sum_{i,j} A_{ij} \overline{x}_i\overline{x}_j|^2 = 0$, due to antisymmetry of $A$. This shows that reduction map is not an exposed map in $M_n(\mathbb{C})$. The only exception is $n=2$. Now, any antisymmtric matrix in $M_2(\mathbb{C})$ is proportional to the Pauli matrix $A = \alpha \sigma_y$ and hence
\begin{equation}\label{}
    A X^{\rm t} A^* = |\alpha|^2 \sigma_y X^{\rm t} \sigma_y = |\alpha|^2 (\mathbb{I}_2 \, {\rm Tr}\, X - X) = |\alpha|^2 R_2(X)\ ,
\end{equation}
which proves that $R_2$ is exposed. The fact that $R_n$ can not be exposed for $n > 2$ is clear since it is not even extreme. Note, however, that $R_n$ is optimal for all $n\geq 2$ \cite{Lew}.
\end{ex}

\begin{ex}[Generalized Choi maps]
Let $\mathcal{H}=\mathcal{K}=\mathbb{C}^3$ and consider the following family of maps in $M_3(\mathbb{C})$

\begin{equation}\label{}
    \phi[a,b,c](X) = \left( \begin{array}{ccc} a x_{11} + bx_{22} + cx_{33} & -x_{12} & -x_{13} \\
    -x_{21} & cx_{11} + a x_{22} + bx_{33}   & -x_{23} \\
    -x_{31} & -x_{32} &  bx_{11} + cx_{22} + ax_{33}   \end{array} \right)\ ,
\end{equation}
with $x_{ij}$ being the matrix elements of $X \in M_3(\mathbb{C})$, and $a,b,c \geq 0$. It was shown \cite{Cho-Kye}
that $\phi[a,b,c]$ is positive (but not completely positive)  if and only if
\begin{enumerate}
\item $0 \leq a <  2\ $,
\item $ a+b+c \geq 2\ $,
\item if $a \leq 1\ $, then $ \ bc \geq (1-a)^2$.
\end{enumerate}
Moreover, being positive it is indecomposable if and only if $ bc < (2-a)^2/4$. Note, that $\phi[1,1,0]$ and $\phi[1,0,1]$ reproduce celebrated Choi map and its dual, whereas $\phi[0,1,1]=R_3$. It was already known that $\phi[1,1,0]$ and $\phi[1,0,1]$ are extreme but not exposed. Recently, it was shown \cite{Filip,Kye-nasza,Gniewko-opt} that a subfamily corresponding to $a=2-b-c$, $a \leq 1$ and $bc=(1-a)^2$ defines optimal maps. Moreover, in a recent paper \cite{Kye-exposed}  it was proved  that maps from this subfamily with $a \in (0,1)$ are exposed. It is the first example of a positive indecomposable exposed map in $M_3(\mathbb{C})$.

\end{ex}

\section{Breuer-Hall maps}

Following \cite{B,H} let us consider $\mathcal{H}=\mathcal{K} = \mathbb{C}^{2n}$ together with a positive map
\begin{equation}\label{BH}
    \phi_{BH}(X) = \mathbb{I}_{2n} \, {\rm Tr} X - X - U X^{\rm t}U^*\ ,
\end{equation}
where $U$ is an arbitrary antisymmetric unitary matrix in $M_{2n}(\mathbb{C})$. It was shown \cite{B} that $\phi_{BH}$ is optimal (even nd-optimal \cite{Lew,B}).

\begin{thm} A map $\phi_{BH}$ is exposed.
\end{thm}

\noindent Proof: one has
\begin{equation}\label{}
    [\phi_{BH}]' = {\rm conv} \{\ P_x \ot P_y \ ; \  \<x|x\>\<y|y\> - |\<y|x\>|^2 - |\<y|U\overline{x}\>|^2 = 0 \ \} \ .
\end{equation}
Let $||x||=||y||=1$. Note that projectors $|x\>\<x|$ and $|U\overline{x}\>\<U\overline{x}|$ are mutually orthogonal. Hence, $|\<y|x\>|^2 + |\<y|U\overline{x}\>|^2 = 1$ if and only if  either $y = e^{i\alpha} x$ or $y = e^{i\alpha} U\overline{x}$ (with $\alpha \in \mathbb{R}$).
It shows that the face $[\phi_{BH}]' = {\rm conv}\{ \, P_x \ot P_x, \, P_x \ot P_{U\overline{x}}\, \}$. Let us characterize maps belonging the double dual  $[\phi_{BH}]''$, that is, maps $\phi$ such that $[\phi]' = [\phi_{BH}]'$. Since $\<x|\phi(P_x)|x\>=0$ any such map has the following form \begin{equation}\label{}
    \phi(X) = \sum_k \lambda_k A_k X^{\rm t} A_k^*\ ,
\end{equation}
where $A_k$ are antisymmetric matrices from $M_{2n}(\mathbb{C})$ and $\lambda_k \in \mathbb{R}$. A set of real numbers $\{ \lambda_k\}$ may be divided into positive $\{ \lambda_k^+\}$ and  negative $\{ \lambda_k^- \}$ elements, respectively. Hence
\begin{equation}\label{}
    \phi(X) = \sum_k \lambda_k^+ A_k X^{\rm t} A_k^* +  \sum_l \lambda_l^- A_l X^{\rm t} A_l^* =
    \sum_k  B_k X^{\rm t} B_k^*  - \sum_l C_l X^{\rm t} C_l^* \ ,
\end{equation}
where $B_k = (\lambda_k^+)^{1/2} A_k$ and $C_l = (-\lambda_l^-)^{1/2} A_l$. Now, using $\<U\overline{x}|\phi(P_x)|U\overline{x}\> = 0$ one obtains
\begin{equation}\label{}
    \sum_k  |\<\overline{x}|U^* B_k |\overline{x}\>|^2  - \sum_l |\<\overline{x}|U^*C_l |\overline{x}\>|^2 = 0\ .
\end{equation}
Since we are interested in the ray $[\phi]$ we can always rescale $A_k$ such that
 \begin{equation}\label{}
  \sum_k  |\<\overline{x}|U^* B_k |\overline{x}\>|^2  = \sum_l |\<\overline{x}|U^*C_l |\overline{x}\>|^2 = 1\ .
\end{equation}
In order to satisfy
\begin{equation}\label{}
\sum_k |\<\overline{x}|U^*D_k |\overline{x}\>|^2 = 1\ ,
\end{equation}
where $D_k$ are antisymmetric matrices ($B_k$ or $C_l$) one has two possibilities:

1)  $D_1 \neq 0$  and $D_2 = D_3 = \ldots = 0$, or

2) a set  $\{D_k\}$ defines a basis in the space of antisymmetric matrices.

\noindent In the first case one has $D_1=U$. In the second case one has the following

\begin{lemma} Let a set  $\{D_k\}$ define a basis in the space of antisymmetric matrices in $M_{2n}(\mathbb{C})$, that is, $D_k \equiv D_{ij}= V(e_{ij} - e_{ji})V^*\ (1\leq i<j \leq 2n)$ with arbitrary unitary matrix $V$, then
\begin{equation}\label{}
    \sum_{i,j} D_{ij} |\overline{x}\>\<\overline{x}| D_{ij}^* = \mathbb{I}_{2n} - |x\>\<x|\ ,
\end{equation}
for any $||x||=1$.
\end{lemma}
Hence, the map $\phi$ reads as follows
\begin{equation}\label{P1}
    \phi(P_x) = (\mathbb{I}_{2n} - |x\>\<x|  ) - U|\overline{x}\>\<\overline{x}|U^*\ ,
\end{equation}
or
\begin{equation}\label{P2}
    \phi(P_x) = U|\overline{x}\>\<\overline{x}|U^* - (\mathbb{I}_{2n} - |x\>\<x|  ) \ .
\end{equation}
Clearly only (\ref{P1}) defines a positive map and it reproduces $\phi_{BH}$ which ends the proof. \hfill $\Box$

\begin{remark} Note, that
\begin{equation}\label{}
    \phi_{BH} = R_{2n} - \phi^U\ ,
\end{equation}
where $R_{2n}$ is the reduction map in $M_{2n}(\mathbb{C})$ and $\phi^U$ is defined in (\ref{V-up}). It shows that subtracting from
the reduction map an exposed map $\phi^U$ one ends up with an exposed map $\phi_{BH}$. It shows that the reduction map is a convex combination of two exposed maps $R_{2n} = \phi_{BH} + \phi^U$ and hence cannot be extreme map. Note, however, that $R_{2n}$ belongs to the face of optimal maps.

\end{remark}

\begin{remark} Taking $n=2$ and $U = \mathbb{I}_2 \ot \sigma_y$, one reconstructs well known Robertson map \cite{Robertson} in $M_4(\mathbb{C})$. Robertson
construction may be nicely described in terms of $R_2$ as follows \cite{atomic}
\begin{equation}\label{R4}
    \phi_{Rob}(X) =  \left( \begin{array}{c|c} \mathbb{I}_2\,
\mbox{Tr} X_{22} &  -[X_{12} + R_2(X_{21})] \\ \hline  -[X_{21} +
R_2(X_{12})] & \mathbb{I}_2\, \mbox{Tr} X_{11}
\end{array} \right) \ ,
\end{equation}
where $X = \sum_{k,l=1}^2 e_{kl} \ot X_{kl}$, with $X_{kl} \in M_2(\mathbb{C})$. It was already proved by Robertson that $\phi_{Rob}$ is extreme. Our result shows that it is exposed as well.

\end{remark}

\section{Conlusions}

We have shown that so called Breuer-Hall positive maps \cite{B,H} are exposed in the convex cone of positive maps in $M_{2n}(\mathbb{C})$ and hence this class defines the most efficient tool for detecting quantum entanglement (any entangled state may be detected by some exposed map (entanglement witness)). As a byproduct we proved that a Robertson map in $M_4(\mathbb{C})$ -- which defines a special case within Breuer-Hall class --  is not only extreme, which was already shown by Robertson, but also exposed. Recently \cite{Justyna} we analyzed various multidimensional generalization of Robertson map. It would be interesting to study whether or not they define exposed maps.

\section*{Acknowledgments}

I thank Gniewko Sarbicki for valuable discussions.

\end{document}